\documentclass[rnote]{aa} 
\usepackage{graphicx}
\usepackage{txfonts}
\begin{document}
\title{A Correlation between Blue Straggler and Binary Fractions in the core of Galactic
Globular Clusters\thanks{Based on ACS observations collected with the Hubble
Space Telescope within the observing program GO 10775.}}


\author{A. Sollima\inst{1},
        B. Lanzoni\inst{1}  
        G. Beccari\inst{1,2},
        F. R. Ferraro\inst{1},
        \and
        F. Fusi Pecci\inst{2}
        }

\offprints{A. Sollima}

\institute{Dipartimento di Astronomia, Universit\`a di Bologna, via Ranzani 1,
Bologna, 40127-I, Italy\\
\email{antonio.sollima@oabo.inaf.it (AS)}
\and
INAF Osservatorio Astronomico di Bologna, via Ranzani 1, Bologna, 40127-I, Italy}

\date{11/01/2008}

 
\abstract
{Blue Stragglers Stars (BSSs) are thought to form in globular clusters by two
main formation channels: $i)$ mergers induced by stellar collisions and $ii)$
coalescence or mass-transfer between companions in binary systems. The detailed study of the
BSS properties is therefore crucial for understanding the binary evolution
mechanisms, and the complex interplay between dynamics and stellar evolution
in dense stellar systems.}
{We present the first comparison between the BSS specific frequency and the
binary fraction in the core of a sample of Galactic globular clusters, with
the aim of investigating the relative efficiency of the two proposed
formation mechanisms.}
{We derived the frequency of BSSs in the core of thirteen low-density Galactic 
globular clusters by using deep ACS@HST observations and
investigated its correlation with the binary fraction and various other
cluster parameters.}
{We observed a correlation between the BSS specific frequency and the
  binary fraction. The significance of the correlation increases by
  including a further dependence on the cluster central velocity
  dispersion.}
{We conclude that the unperturbed evolution of primordial
    binaries could be the dominant BSS formation process, at least in
  low-density environments.}


\authorrunning
\titlerunning 
\maketitle
%

\section{Introduction}
Blue Straggler Stars (BSSs) are objects that, in the color-magnitude diagram
(CMD) of evolved stellar populations, lie along an extension of the Main
Sequence (MS), in a region which is brighter and bluer than the Turn-off
(TO). First discovered by Sandage (1953) in M3, they have been observed in
all Galactic globular clusters (GCs; Piotto et al. 2004), in the field population (Carney et al. 2005), and in
dwarf galaxies of the local group (Momany et al. 2007). 

Their location in
the CMD suggests that BSSs have masses of $1.2 \div 1.5~M_{\odot}$,
significantly larger than those of normal stars in old stellar systems (like
GCs). Thus, they are thought to have increased their mass during their
evolution. Two mechanisms have been proposed for their formation: $i)$
the merger of two stars induced by stellar collision (COL-BSSs; Hills \& Day
1976) and $ii)$ coalescence or mass-transfer between two companions in a binary system
(MT-BSSs; McCrea 1964). The two formation channels are thought to act with
different efficiencies according to the cluster structural parameters (Fusi
Pecci et al. 1992) and they can work simultaneously within the same cluster in
different radial regions, corresponding to widely different stellar densities
(Ferraro et al. 1997; Mapelli et al. 2006). Indeed, collisions are more
frequent in the central region of GCs, because of the high stellar density,
while MT-BSSs mainly populate the cluster periphery, where binary systems can
more easily evolve in isolation without suffering exchange or ionization due
to gravitational encounters. The whole scenario is further complicated by
the cluster dynamical evolution that leads massive systems (like binaries and
BSSs) to sink toward the cluster center in a time-scale comparable to the
cluster relaxation time. 

A possible tool for distinguishing COL-BSSs from MT-BSSs is based on
high-resolution spectroscopic analysis. In fact, anomalous chemical
abundances are expected at the surface of BSSs resulting from
mass-transfer activity (Sarna \& de Greve 1996), while they are not
predicted for COL-BSSs (Lombardi et al. 1995). However, such studies
have just become feasible and they are limited to only a small number
of BSSs in just one cluster (47 Tucanae; Ferraro et al. 2006).

As an alternative way for getting insights on the relative efficiency of the
two formation mechanisms, here we investigate possible correlations between
the BSS population and the host cluster properties. 

\section{Observations and Data Reduction}
\label{obs}
The photometric database considered here is the same used in Sollima
et al. (2007; hereafter S07) for measuring the frequency of binary
systems in the core of thirteen low-density Galactic GCs. It consists
of a set of high-resolution images, obtained with the ACS on board
HST, through the F606W ($V_{606}$) and F814W ($I_{814}$) filters.  As
described in S07, the target clusters are characterized by high
Galactic latitude ($b>15^\circ$), low reddening ($E(B-V)<0.1$) and low
projected central density ($\log\rho'_0<5$, where $\rho'_0$ is in
units of $M_\odot~{\rm arcmin}^{-2}$). These selection criteria have
been chosen in order to limit the effects of field contamination,
differential reddening and crowding on the determination of the binary
fractions. The selected target clusters and their main physical
parameters are listed in Table \ref{tab:sample}. The central
luminosity density $\rho_0$ and the absolute $V$ band magnitude $M_V$
are from Djorgovski (1993), the age $t_{9}$ is from Salaris \& Weiss
(2002)\footnote{Since NGC~6981 is not included in the list of Salaris
  \& Weiss (2002), we converted the ages measured by De Angeli et
  al. (2005) into the Salaris \& Weiss (2002) scale.}, and the
projected central velocity dispersion $\sigma_{\rm v}$ is from
McLaughlin \& Van der Marel (2005).  A detailed description of the
data reduction and calibration procedure can be found in S07 together
with the resulting CMDs for all the target clusters.

As discussed in S07, the binary sequence is well defined and
distinguishable from the MS and the fraction of binary systems
($\xi_{bin}$) found within the cluster core radius ranges from about
10 to $50\%$ (the value of $\xi_{bin}$ are listed in Table
\ref{tab:sample}). Moreover, a number of BSSs populating the bright
part of the CMD is also evident in the data sample. In the following
section we describe the adopted procedure to derive the BSS
frequencies in the core of the target clusters.

\section{BSS population selection}
\label{bss}

\begin{figure}
\centering
\includegraphics[width=8.7cm]{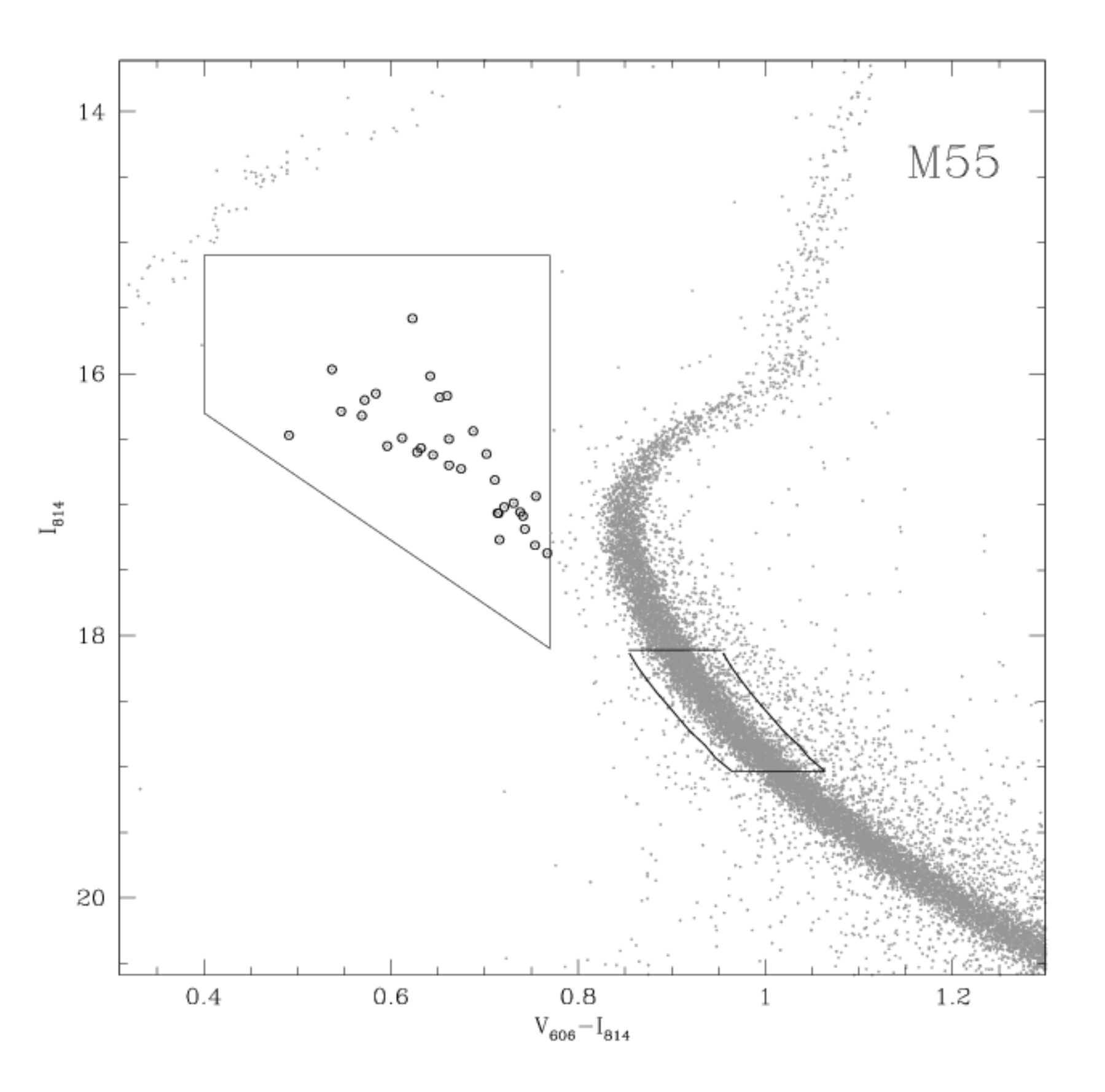}
\caption{($I_{814},~V_{606}-I_{814}$) CMD of M55. The adopted
  selection boxes for the BSS and MS populations are shown. The
  fiducial BSSs are marked with open circles.  }
\label{fig:box} 
\end{figure}

Our primary criterion for the definition of the BSS sample is based on
the location of stars in the ($I_{814},~V_{606}-I_{814}$) CMD. Since
the considered GCs have different distances, reddening, ages and
metallicities, in order to define a homogeneous criterion for the BSS
selection, we shifted in magnitude and color the CMD of each cluster
to match that of M55 (see also Ferraro et al. 1995). For minimizing
the contamination from sub-giant branch, horizontal branch (HB) and
field stars, we limited the analysis to colors bluer than
$(V_{606}-I_{814})<0.47$ and magnitudes in the range $15.1
<I_{814}<18.1$. The adopted BSS selection box (the same for all
  clusters) is shown in Figure \ref{fig:box} and the number of
selected BSSs ($N_{\rm BSS}$) in each target cluster is reported in
Table \ref{tab:sample}.

In order to compare the frequency of BSSs in different GCs, the size
of the total cluster population must be taken into consideration.
Commonly, the number of BSSs is normalized to the number of stars
belonging to a given evolutive sequence (usually the HB, or the Red
Giant Branch, RGB; Ferraro et al. 1995; Piotto et al.  2004; Leigh
et al. 2007).  Unfortunately, in some clusters of our sample the
number of HB and RGB stars is very small, which would make the
derived BSS frequency highly uncertain. We therefore normalized
the number of selected BSSs, to the number of MS stars ($N_{\rm MS}$)
in the magnitude range $18.1<I_{814}<19.1$ and with colors comprised
within $\Delta (V_{606}-I_{814})<0.05$ from the cluster MS mean ridge
line.  Stars in this magnitude range are bright enough to have the
same completeness level of BSSs ($\phi>95\%$) and cover a similar mass
range in all clusters: following the theoretical stellar isochrones by
Cariulo et al. (2004) and taking into account the differences in age
and metallicities among the selected GCs, the TO mass varies between
$0.8~M_{\odot}<M<0.9~M_{\odot}$ .  Hence, we considered $N_{MS}$ a
good indicator of the size of the sampled cluster population.

The possible effect of background and foreground field star
contamination on the BSS and MS selections has been estimated by using
the Galaxy model of Robin et al. (2003). For this purpose a synthetic
catalog covering an area of 0.5 square degree centered on each cluster
(with the centers taken from Djorgovski \& Meylan 1993) has been
retrieved, both in the $V$ and the $I$ Johnson-Cousin bands. A
sub-sample of stars scaled to the size of the ACS field of view
($202\arcsec\times~202\arcsec$) has been randomly extracted, and the
Johnson-Cousin magnitudes have been converted into the ACS photometric
system by using the transformations of Sirianni et al. (2005).
The number of contaminating field stars turns out to be neglegible ($<$2) 
in all the clusters of our sample.
Then, we counted the number of field stars included in the BSS and MS
selection boxes ($N_{field}^{\rm BSS}$ and $N_{field}^{\rm MS}$,
respectively), and we computed the BSS specific frequency as:

\begin{equation}
F = \frac{N_{\rm BSS}-N_{field}^{\rm BSS}}{N_{\rm MS}-N_{field}^{\rm MS}}
\end{equation}

The resulting values for the thirteen target clusters are listed in Table
\ref{tab:sample}.  In order to check the suitability of the adopted
normalization criterion, we compared the derived BSS specific
frequencies with those measured by Leigh et al. (2007, and
uptdated by Sills 2008, private communication) for the four
clusters in common with our sample (namely, NGC~4590, NGC~6362,
NGC~6723 and NGC~6981). These authors normalized the number of BSSs in
the cluster cores to that of red giant branch stars in the same
magnitude range. In spite of the different normalization criterion,
the specific frequencies calculated for the four clusters in common
correlate very well (with a correlation coefficient $r=0.92$).

%
\begin{table}
\begin{minipage}[t]{\columnwidth}
\caption{Main physical parameters of the target globular clusters}
\label{tab:sample}
\centering
\renewcommand{\footnoterule}{}  
\begin{tabular}{lccccccr}
\hline \hline
 Name & $\log~\rho_0$     & $M_V$ & $t_9$ & $\sigma_{\rm v}$ & $\xi_{bin}$ & $N_{BSS}$ & $F$\\
      & $L_\odot~pc^{-3}$ &       & Gyr   &  Km s$^{-1}$     & \%          &           & \% \\
\hline
 NGC 288    & 1.80  & -6.63 &	 11.3  & 2.79 & 11.6 & 43 & 1.9\\
 NGC 4590   & 2.52  & -7.73 &	 11.2  & 3.59 & 14.2 & 62 & 1.5\\
 NGC 5053   & 0.51  & -7.07 &	 10.8  & 1.79 & 11.0 & 15 & 1.0\\
 NGC 5466   & 0.68  & -6.83 &	 12.2  & 1.86 &  9.5 & 39 & 1.7\\
 NGC 5897   & 1.32  & -7.27 &	 12.3  & 2.94 & 13.2 & 53 & 1.3\\
 NGC 6101   & 1.57  & -6.82 &	 10.7  & 2.91 & 15.6 & 60 & 1.1\\
 NGC 6362   & 2.23  & -6.69 &	 11.0  & 3.83 & 11.8 & 42 & 1.5\\
 NGC 6723   & 2.71  & -7.95 &	 11.6  & 5.52 & 16.1 & 35 & 0.5\\
 NGC 6981   & 2.26  & -7.16 &  	  9.5 & 3.66  & 28.1 & 65 & 1.5 \\ 
 M55        & 2.12 & -7.47 & 	12.3  & 3.73  &  9.6 & 32 & 0.8\\
 Arp 2      & -0.35& -5.30 & 	9.7   & 0.72  & 32.9 & 33 & 2.3\\
 Terzan 7   & 1.97 & -5.99 & 	7.4   & 2.55  & 50.9 & 33 & 3.4\\
 Palomar 12 & 0.68 & -4.78 & 	6.4   & 0.76  & 40.8 &  9 & 2.5\\
\hline
\end{tabular}
\end{minipage}
\end{table}
%

\section{Correlations between BSS specific frequency and cluster parameters}
\label{correl}

With the aim of shading light on the BSS formation mechanisms, we
correlated the BSS specific frequency ($F$) calculated above with the
core binary fraction ($\xi_{bin}$) obtained by S07 and with various
other physical parameters of the thirteen GCs of the sample. In
particular, we searched for possible correlations with the cluster
central density, total magnitude, age, and central velocity dispersion
(see Table \ref{tab:sample}), as well as with the stellar collision
rate (defined as $\Gamma \equiv \rho_0^2~r_c^3/\sigma_{\rm v}$; Pooley
\& Hut 2006), the global metallicity ([M/H], from Ferraro et
al. 1999\footnote{For NGC~6101, NGC~6362, NGC~6723 and Palomar 12 not
  included in the list of Ferraro et al. (1999) we transformed the
  metallicity [Fe/H] from Zinn \& West (1984) into the global
  metallicity [M/H] following the prescriptions of Ferraro et
  al. (1999).}), the cluster evaporation rate ($\nu$, from Gnedin \&
Ostriker 1997), concentration ($c$, from Trager et al. 1995),
half-mass relaxation time ($t_{rh}$) and half-mass radius ($r_h$, from
Djorgovski 1993).

\begin{figure}
\centering
\includegraphics[width=8.7cm]{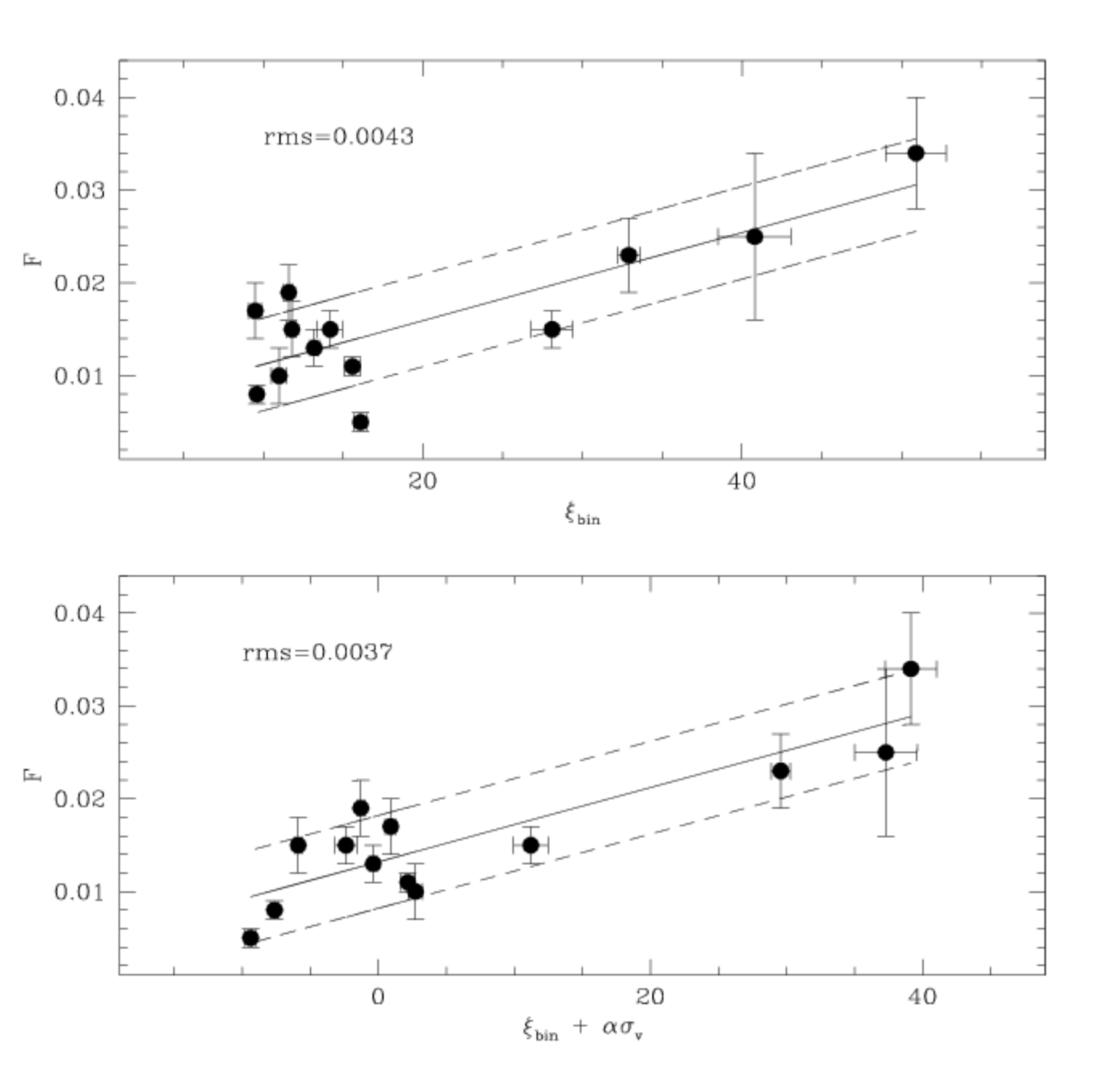}
\caption{Specific frequency of BSSs as a function of the core binary fraction 
($upper~panel$) and of the best-fit linear
combination of core binary fraction and cluster velocity dispersion 
($bottom~panel$, obtained with $\alpha=-4.62$).
Solid lines indicate the best-fit lines; dashed lines mark the boundaries of 
the $\pm$0.005 range in F with respect to the best-fit lines.}
\label{fig:bssbinsig}
\end{figure}

The Pearson's linear correlation coefficients and confidence levels
for the considered parameter pairs are listed in Table
\ref{tab:spear}.  The strongest correlation is found between the BSS
specific frequency and the binary fraction.  A noticeable correlation
with the absolute magnitude and anticorrelations with the cluster age
and central velocity dispersion are also apparent.  Considering that
the age estimates are quite uncertain and span a narrow range of
values in our sample, any firm conclusion about the possible
dependence of $F$ on the cluster age is premature at the
moment. However, if confirmed, such an anticorrelation might be a
consequence of the analogous anticorrelation between binary fraction
and age already discussed by S07, and would suggest that binary
disruption processes in the core of GCs become more efficient with
time, thus reducing the fraction of both binaries and BSS in the core
of the oldest clusters (as suggested by Ivanova et al. 2005, but at
odds with Hurley et al. 2007). In any case, even if the
anticorrelation between $F$ (or $\xi_{bin}$) and $t_9$ is not
confirmed, the correlation between the BSS specific frequency and the
binary fraction still holds. The latter is shown in the upper
panel of Figure \ref{fig:bssbinsig} and it suggests that the main
formation channel of BSS in these clusters is the unperturbed
evolution of primordial binary systems.

We stress that, since the considered parameters do not constitute an
orthonormal basis, most of the less significant correlations could be
artificially induced by primary correlations among other
parameters. In order to investigate the possible dependence of $F$ on
more than one cluster parameter, we applied the {\it Bayesian
  Information Criterion (BIC)} test (Schwarz 1978) to our dataset. We
assumed the BSS specific frequency $F$ to be a linear combination of a
sub-sample of $p$ parameters ($\lambda_i$) selected among those listed
above:

\begin{equation}
F_{p}^{fit}=\sum_{i=1}^{p} \alpha_i~\lambda_i + \alpha_{p+1}
\end{equation}
and we calculated the quantity

\begin{equation}
BIC=\log~ L_p -\frac{p}{2} \log~N, 
\end{equation}
where $L_p$ is the likelihood and $N$ is the dimension of the sample ($N=13$
in our case). The likelihood is calculated as $\log~ L_p =
\sum_{j=1}^{N}\log~ P_{p,j}$, where
\begin{equation}
P_{p,j} =\frac{e^{\frac{-\left(F_j-F_{p,j}^{fit}\right)^2}{2\sigma_F^2}}}{\sigma_F\sqrt{2\pi}} 
\end{equation}
with $F_j$ being the BSS fraction of cluster $j$, and $\sigma_F$ being
the residual of the fit. The $p$ parameters that maximize the quantity
$BIC$ are the most probable correlators with $F$. For our dataset the
maximum value of $BIC$ is obtained with $p=2$, $\lambda_1=\xi_{bin}$
and $\lambda_2 = \sigma_{\rm v}$. The correlation between $F$,
$\xi_{bin}$ and $\sigma_{\rm v}$ is shown in Figure
\ref{fig:bssbinsig}, together with the best fit line. All the other
3-variate or higher-order correlations turn out to be
non-significant.

A note of caution is worth: as can be seen, these correlations
are essentially due to the four clusters (namely Terzan~7,
Palomar~12, Arp~2 and NGC~6981) with the largest ($>20\%$) binary
fraction, while the other GCs in our sample define a group with an
average binary content $\xi_{bin}\sim 10\%$ and a BSS frequency $F\sim
1.2\%$. In particular, all the correlations that we have found are
mainly driven by Terzan~7, Palomar~12 and Arp~2, which are the GCs not
only with the largest binary and BSS fractions, but also with the
lowest absolute magnitudes and the lowest ages.  These three clusters
are also the most distant from the Sun and they are thought to belong
to the Sagittarius Stream (Bellazzini et. al 2003). Thus, they might
be stellar systems with intrinsically different origins and
properties, and the correlations found here might not be appropriate
for the overall class of "genuine" Galactic GCs
(GGCs).
By excluding these three clusters from the sample, the anticorrelation
with the total luminosity still holds (even if with lower
significance), in agreement with what already found by Piotto et
al. (2004), Sandquist (2005) and Leigh et al. (2007) for larger
samples of GCs with different characteristics, and for BSS populations
selected and normalized with different criteria. Instead, the
anticorrelation with the age disappears, thus bringing our results in
agreement with those of Leigh et al. (2007; see also De Marchi et
al. 2006).  The correlation between the BSS and the binary
fractions also disappears excluding the three "Sagittarius"
clusters (SgrGCs), with the range in $\xi_{bin}$ spanned by the
remaining sample being very small.  Note however that both the
average BSS and binary frequencies are higher in SgrGCs 
($\langle F_{\rm BSS}\rangle=2.7\%$ and $\langle \xi_{bin}\rangle =41.5\%$) 
than in GGCs ($\langle F_{\rm BSS}\rangle=1.3\%$ and 
$\langle \xi_{bin}\rangle= 14\%$). Thus,
even assuming that SgrGCs were born with different initial conditions from 
GGCs, the observational facts presented here allow to
draw (at least) the "conservative" conclusion that binary-rich enviroments
tend to produce more BSS in the low-density cluster regime (where
collisions are expected to play a minor role in the 
formation/destruction of binaries).

%
\begin{table}
\begin{minipage}[t]{\columnwidth}
\caption{Pearson's correlation coefficients ($r$) and confidence levels (P)}
\label{tab:spear}
\centering
\renewcommand{\footnoterule}{}  
\begin{tabular}{lcc}
\hline \hline
  Parameter     & $r$    & P \\
\hline
$\xi_{bin}$     & 0.820  & 0.9998\\
$M_V$           & 0.774  & 0.9985\\
$t_9$           & -0.758 & 0.998\\
$\sigma_{\rm v}$& -0.594 & 0.98\\
$[M/H]$         & 0.403  & 0.92\\
$\rho_0$        & -0.311 & 0.85\\
$r_h$           & 0.299  & 0.84\\
$\Gamma$      & -0.158 & $<0.75$\\
$c$             & 0.129  & $<0.75$\\
$\nu$           & 0.099  & $<0.75$\\
$\log~t_{rh}$  & 0.020  & $<0.75$\\
\hline
\end{tabular}
\end{minipage}
\end{table}
%
%
%

\section{Conclusions}
We measured the BSS specific frequency in the core of thirteen
low-density Galactic GCs and investigated its correlation with
different dynamical and general cluster parameters. We found 
evidences that, at least in this density regime, binary-rich
environments are more efficient in producing BSS. No correlations
have been found with the cluster central density, concentration,
stellar collision rate, and half-mass relaxation time, in agreement
with the results of Piotto et al. (2004) and Leigh et
al. (2007). These evidences indicate that the collisional channel for
the BSS formation has a very small efficiency in low-density GCs,
while the mechanisms involving the unperturbed evolution of binary
systems are dominant.

The higher significance of the trivariate correlation among the BSS frequency,
the binary fractions and the cluster velocity dispersion, indicates that,
for a given binary fraction, the BSS specific frequency decreases with
increasing velocity dispersion. This finding might be connected with the
effect of the cluster velocity dispersion in the dynamical evolution of
binary systems. In fact, a small cluster velocity dispersion corresponds to a
lower energy limit between soft and hard binaries\footnote{A binary is defined
{\it soft} ({\it hard}) if its binding energy ($E = -G~m_{1}~m_{2}/2~a$, with $a$ 
being the orbital separation of the two components) is smaller (larger) than
the mean kinetic energy of normal cluster stars ($K = m~\sigma_{\rm v}^2$, 
with $m$ being the average mass of cluster stars).}, 
i.e., to a larger fraction of hard binary systems. 
Since the natural evolution of hard binaries is to increase 
their binding energy (i.e. decrease their orbital separation; Heggie 1975), this implies 
that low velocity dispersion GCs should host a larger fraction of hard (and close)
binaries, able to both survive possible stellar encounters, and activate 
mass-transfer and/or merging processes between the companions. 
A larger fraction of BSSs formed by the evolution of
primordial binaries is therefore expected in lower velocity dispersion GCs (see also Davies et al. 2004).
Such an effect of $\sigma_{\rm v}$ (in terms of both hardening and shrinking
the binary systems) might be, in turn, at the origin of the inverse
correlation between the BSS frequency and the cluster total luminosity (mass)
observed in open clusters (De Marchi et al. 2006), low density GCs (Sandquist
2005), as well as high density GCs (Piotto et al. 2004; Leigh et al. 2007).
Indeed, the more massive GCs have larger central velocity dispersions (as a
consequence of the virial theorem for systems with similar radii, as GCs; see
Fig. 1 of Djorgovski 1995).  For most of these clusters, however, the binary
fraction is still unknown and the trivariate correlation between $F$,
$\xi_{bin}$ and $\sigma_{\rm v}$ cannot be derived. If its significance and
its interpretation in terms of the velocity dispersion effect is confirmed 
also in high-density clusters, then stellar collisions might play a secondary
role in the production of BSSs, and the evolution of primordial binaries
should always be the dominant process. For the moment, however, these
conclusions remain speculative, since the sample of GCs analyzed here, in
spite of being the largest to date with known binary fraction, is still too
small for statistically reliable assessments. Enlarging the sample of GCs
with known binary and BSS fraction is therefore essential and urgent to
verify the findings presented in this paper on a more robust statistical
basis.

\begin{acknowledgements}
This research was supported by the Ministero dell'Istruzione, Universit\`a e
Ricerca and the Agenzia Spaziale Italiana. 
This research is part of the {\it Progetti strategici di Ateneo 2006} granted 
by the Bologna University. We warmly thank Paolo Montegriffo
for assistance during catalogs cross-correlation.
We also thank Alison Sills, the referee of our paper, for her helpful comments
and suggestions.
\end{acknowledgements}

\end{document}